\documentclass{article}
\usepackage{amscd,amsmath,amssymb,bbm}
\usepackage{amsfonts}
\newcommand{\p}[1]{(\ref{#1})}

\newcommand{\bZ}{{\overline{Z}}{}}

\newcommand{\bD}{\overline{D}{}}
\newcommand{\bR}{\overline{R}{}}

\newcommand{\bpsi}{{\bar\psi}}

\newcommand{\bp}{{\bar p}}
\newcommand{\bz}{{\bar z}}

\newcommand{\be}{\begin{equation}}
\newcommand{\ee}{\end{equation}}
\newcommand{\bea}{\begin{eqnarray}}
\newcommand{\eea}{\end{eqnarray}}
\newcommand{\ba}{\begin{array}}
\newcommand{\ea}{\end{array}}

\newcommand{\nn}{\nonumber}

\def\im{{\rm i}}

\def\={\ =\ }

\def\Nf{$\cal N${=\,}4~}

\begin{document}
\thispagestyle{empty}
\vspace{2cm}
\begin{flushright}
\end{flushright}\vspace{2cm}
\begin{center}
{\LARGE\bf $CP^n$ supersymmetric mechanics in $U(n)$ background gauge
 fields}
\end{center}
\vspace{1cm}

\begin{center}
{\Large
Stefano Bellucci$\,{}^{a}$, Sergey Krivonos$\,{}^{b}$  and
Anton Sutulin$\,{}^{b}$
}\\
\vspace{1.0cm}
${}^a$ {\it
INFN-Laboratori Nazionali di Frascati,
Via E. Fermi 40, 00044 Frascati, Italy} \vspace{0.2cm}

${}^b$
{\it Bogoliubov  Laboratory of Theoretical Physics,
JINR, 141980 Dubna, Russia}
\vspace{0.2cm}
\end{center}
\vspace{3cm}

\begin{abstract}
\noindent We construct a new \Nf supersymmetric mechanics describing the motion of a particle
over a $CP^n$ manifold in $U(n)$ background gauge fields.

\end{abstract}

\newpage
\setcounter{page}{1}

\setcounter{equation}0
\section{Introduction}
The study of quantum Hall effect in higher (larger then two) dimensions has been of some interest
in the last several years, following the analysis by Zhang and Hu \cite{ZH}.
They considered the Landau problem for charged fermions on $S^4$ with a background
magnetic field which is $SU(2)$ instanton. A number of papers have extended the original idea
in many aspects (see e.g. \cite{KN1} and references therein).
One of such extensions concerns the analysis of quantum Hall effect on complex projective spaces
$CP^n$ \cite{Kar1, Nair1}. The corresponding bulk and edge actions were derived \cite{Kar1}.
In addition, in \cite{Nair1} it has been shown that the bulk contribution coincides
with the Chern-Simons action.

The geometry of the $CP^n$ space is quite simple - this is just the coset $SU(n+1)/U(n)$.
For any $G/H$ coset one has at hands the analogue of a constant background field --
the $H$-valued connection on $G/H$.
Thus, the case of $CP^n$ allows for both Abelian and non-Abelian background fields \cite{KNair}.
Moreover, the system describing the motion of particles over the $CP^n$ manifold could
be easily extended to possess \Nf supersymmetry \cite{sm1, BN1, BN2} in the absence
of background fields. Thus, it seems to be a proper task to include the coupling with
the background gauge fields in the \Nf supersymmetric system on $CP^n$.
That is just what we are doing in the present paper.
We explicitly construct the \Nf supercharges and Hamiltonian which describe the motion
of a particle over the $CP^n$ manifold in the presence of background $U(n)$  fields.
The corresponding gauge potential is proportional to the $U(n)$-connection on $SU(n+1)/U(n)$.
Surprisingly, this form of the gauge potential is dictated by  \Nf supersymmetry.
It turns out that  \Nf supersymmetry demands the presence of additional pure potential
terms in the Hamiltonian.
In the simplest case of the $CP^1$ system this potential is just a harmonic oscillator one.

The paper is organized as follows. In section 2 we review the \Nf supersymmetric mechanics
on $CP^n$. The supercharges and the Hamiltonian are derived in section 3.
We conclude our work with a short discussion.

\setcounter{equation}0
\section{$CP^n$ mechanics with \Nf supersymmetry}
The construction of \Nf supersymmetric mechanics on $CP^n$ manifold\footnote{Let us stress that $CP^n$ is a geometry of target space, while all fields depend on time $t$ only.} is almost trivial.
Indeed, if we  take $n$ complex \Nf chiral superfields $Z^\alpha(\theta_i, \bar\theta{}^i), \bZ_\alpha(\theta_i, \bar\theta{}^i)$ defined in $d=1, N=4$ superspace ${\mathbb R}^{(1,4)}=\left\{t,\theta_i, \bar\theta{}^i\right\}, i=1,2$ and obeying conditions
\be\label{eq1}
D^i Z^\alpha=0 ,\quad \bD_i \bZ_\alpha=0 , \qquad \alpha=1\ldots n,
\ee
where \Nf covariant derivatives are defined as
\be\label{D}
\left\{ D^i , \bD_j \right\} = 2 \im \delta^i_j \partial_t,
\ee
then the superfields action $S$
\be\label{actionSF}
S= \int dt\, d^4 \theta  \; \log\left[ 1 + Z^\alpha \bZ_\alpha \right]
\ee
does all the job, completely defining the model.
The explicit form of Lagrangian density in \p{actionSF} immediately follows from invariance
of the action with respect to the $SU(n+1)$ group, which is realized on the superfields $Z,\bZ$ as
\be\label{su0}
\delta Z^\alpha = a^\alpha + Z^\alpha \left( Z^\beta {\bar a}_\beta\right),\quad
\delta \bZ_\alpha = {\bar a}_\alpha + \bZ_\alpha \left(a^\beta \bZ_\beta \right),
\ee where $a^\alpha, {\bar a}_\alpha$ are the parameters of the
coset $SU(n+1)/U(n)$ transformations.

If we instead will not fix the integrand in the action $S$ \p{actionSF} leaving it to be
an arbitrary function ${\cal L}(Z,\bZ)$, then the resulting system will describe
a supersymmetric mechanics on an arbitrary $n$-dimensional K\"{a}hler manifold
(see e.g. \cite{sm1,BN1,BN2}). In the case of one superfield $Z,\bZ$ such a system
has been firstly constructed in \cite{BP}. Recently, supersymmetric mechanics on complex
manifolds has been considered in \cite{IS1}.

To fix our notations and for completeness, let us shortly discuss the Hamiltonian description
of the \Nf supersymmetric $CP^n$ mechanics which directly follows from \p{actionSF}
after passing to the components and removing the auxiliary fields.

So, our basic ingredients are bosonic variables $\left\{ z^\alpha, {\bar z}_\alpha\right\}$
which parameterize the coset $SU(n+1)/U(n)$ and fermionic variables
$\left\{\psi^\alpha_i,\bpsi_{\alpha}^i\right\}$:
\be\label{comp1}
z^\alpha = Z^\alpha|,\quad {\bar z}_\alpha=\bZ_\alpha|, \qquad
\psi^\alpha_i=\bD_i Z^\alpha| , \quad \bpsi_{\alpha}^i= D^i \bZ_\alpha|,
\ee
where $(\ldots)|$ denotes $\theta_i=\bar\theta{}^i=0$ limit.
In what follows we will pay a great attention to $U(n)$ properties of our model.
That is why we decided to keep the corresponding indices $\alpha,\beta$ of our fields \p{comp1}
in a proper position. For the $SU(n+1)$ group we will fix the commutation relations to be
\bea\label{sun1}
&& i\left[ R_\alpha, \bR^\beta\right] = J_\alpha{}^\beta, \quad
i\left[J_\alpha{}^\beta, J_\gamma{}^\sigma \right]=
\delta_\gamma^\beta J_\alpha{}^\sigma - \delta_\alpha^\sigma J_\gamma{}^\beta, \nn \\
&& i\left[ J_\alpha{}^\beta,  R_\gamma \right] =\delta_\gamma^\beta R_\alpha+\delta_\alpha^\beta R_\gamma, \quad
i\left[ J_\alpha{}^\beta, \bR^\gamma \right] = -\delta_\alpha^\gamma \bR^\beta-\delta_\alpha^\beta \bR^\gamma .
\eea
Thus, the generators $ R_\alpha, \bR^\alpha$ belong to the coset $SU(n+1)/U(n)$,
while the $J_\alpha{}^\beta$ form $U(n)$. In addition we choose these generators to be
anti-hermitian ones
\be\label{anti}
\left( R_\alpha \right)^\dagger = -\bR^\alpha, \quad
\left( J_\alpha{}^\beta\right)^\dagger= -J_\beta{}^\alpha.
\ee
After introducing the momenta for all our variables and passing to Dirac brackets
we will obtain the following set of relations
\footnote{As usual, the bosonic momenta are shifted by
$\psi\cdot \bpsi$ terms in this basis.}
\bea\label{PB}
&& \left\{ \psi^\alpha_i, \bpsi_\beta^j\right\} =i \delta_i^j \left( g^{-1}\right)_\beta{}^\alpha, \qquad
\left\{ p_\alpha , \bp^\beta\right\}=
-i \left( g_\alpha{}^\beta g_\mu{}^\nu+g_\alpha{}^\nu g_\mu{}^\beta \right) \bpsi_\nu^i \psi_i^\mu, \nn \\
&& \left\{ p_\alpha, \psi_i^\beta\right\} =-\frac{1}{\left( 1+ z\cdot \bz\right)}
\left[ \bz_\alpha \psi_i^\beta +\delta_\alpha^\beta \,\psi_i^\gamma \bz_\gamma\right], \quad
\left\{ \bp^\alpha, \bpsi^i_\beta\right\} =-\frac{1}{\left( 1+ z\cdot \bz\right)}
\left[ z^\alpha \bpsi^i_\beta +\delta^\alpha_\beta \, z^\gamma \bpsi^i_\gamma \right].
\eea
Here, the $CP^n$ metric $g_\alpha{}^\beta$ has the standard Fubini-Study form
\be\label{metrics}
g_\alpha{}^\beta=\frac{1}{\left( 1+ z\cdot \bz\right)}
\left[\delta_\alpha^\beta -\frac{\bz_\alpha z^\beta}{\left( 1+ z\cdot \bz\right)}\right], \quad
z\cdot \bz\equiv z^\alpha \bz_\alpha.
\ee

Now, it is not too hard to check that the supercharges $Q^i, {\overline Q}_i$ have the extremely
simple form \cite{sm1,BN1,BN2}
\be\label{superch}
Q^i = \bp^\alpha \, \bpsi^i_\alpha, \qquad
{\overline Q}_i = \psi^\alpha_i\, p_\alpha.
\ee
They are perfectly anticommuting (in virtue of \p{PB}) as
\be\label{Q1}
\left\{ Q^i, {\overline Q}_j\right\}= i\delta^i_j H, \qquad
\left\{ Q^i, Q^j\right\} = \left\{{\overline Q}_i, {\overline Q}_j\right\}=0,
\ee
where the Hamiltonian $H$ reads
\footnote{The $su(2)$ indices are raised and
lowered as $A_i=\varepsilon_{ij}A^j, A^i=\varepsilon^{ij} A_j$
with $\varepsilon_{12}=\varepsilon^{21}=1$.}
\be\label{H}
H=\bp^\alpha\, \left( g^{-1}\right)_\alpha{}^\beta \, p_\beta +
\frac{1}{4} \left( g_\mu{}^\alpha g_\rho{}^\sigma+ g_\mu{}^\sigma
g_\rho{}^\alpha \right) \bpsi_{\alpha\, i}\bpsi^i_\sigma\, \psi^{\rho\,j}\psi^\mu_j.
\ee
In principle, one may modify the supercharges and Hamiltonian by including potential
terms \cite{BN1,BN2}, but here we will be interested in including the interaction
with non-Abelian background fields which looks in itself rather complicated.
Therefore we will ignore such possible modifications in what follows.

Finally, we will need the explicit expressions for the vielbeins $e_\alpha{}^\beta$ and
$U(n)$-connections $\omega_\alpha{}^\beta$ on the $CP^n$ manifold, which we choose as \cite{klw}
\bea
&&
e_\alpha{}^\beta = \frac{1}{\sqrt{1+z\cdot \bz}}\left[ \delta_\alpha^\beta -
\frac{\bz_\alpha z^\beta}{\sqrt{1+z\cdot \bz}\left(1+\sqrt{1+z\cdot \bz}\right)}\right],
\label{vb} \\
&&
\omega_\alpha{}^\beta= \frac{1}{\sqrt{ 1+ z \cdot \bz
}\left(1+\sqrt{1+ z \cdot \bz}\right)}\left[ \delta_\alpha^\beta-
\frac{\bz_\alpha z^\beta}{2\,\sqrt{1+ z \cdot \bz}\left(1+\sqrt{1+ z \cdot \bz}\right)}\right].
\label{con}
\eea
With our definition of the $SU(n+1)$ algebra \p{sun1}, these quantities enter the standard Cartan forms as
\be\label{CF}
g^{-1}\,d g = dz^\alpha\, e_\alpha{}^\beta R_\beta +\bR^\alpha e_\alpha{}^\beta d\bz_\beta
+ i J_\alpha{}^\beta \left( z^\alpha \, \omega_\beta{}^\gamma d \bz_\gamma
- dz^\gamma\, \omega_\gamma{}^\alpha \bz_\beta \right),
\ee
where
\be\label{gsu}
g=e^{x^\alpha\, R_\alpha +{\bar x}_\alpha \bR^\alpha}, \quad
\mbox{ and } \quad z^\alpha \equiv \frac{ \tan \sqrt{ x \cdot {\bar x}}}{\sqrt{ x \cdot {\bar x}}}x^\alpha .
\ee

\setcounter{equation}0
\section{Gauge fields: construction}
It is curious, but the simplest form of the supercharges \p{superch} does not help
in the coupling with background gauge fields. One may easily check that the standard coupling
by shifting bosonic momenta in supercharges does not properly work. Our idea is to introduce
the coupling simultaneously with all currents of the $SU(n+1)$ and/or $SU(1,n)$ groups.
Thus, let us introduce the isospin currents spanning $SU(n+1)$ and/or $SU(1,n)$, respectively
\bea\label{su}
&& \left\{ R_\alpha, \bR^\beta\right\} = -A  J_\alpha{}^\beta, \quad
\left\{ J_\alpha{}^\beta, J_\gamma{}^\sigma \right\}
=\delta_\gamma^\beta J_\alpha{}^\sigma- \delta_\alpha^\sigma J_\gamma{}^\beta, \nn \\
&& \left\{ J_\alpha{}^\beta,  R_\gamma \right\} =\delta_\gamma^\beta R_\alpha+\delta_\alpha^\beta R_\gamma, \quad
\left\{ J_\alpha{}^\beta, \bR^\gamma \right\} =-\delta_\alpha^\gamma \bR^\beta-\delta_\alpha^\beta \bR^\gamma .
\eea
The coefficient $A=\pm 1$ in the first line corresponds to the choice of $SU(1,n)$ or $SU(n+1)$, respectively.
It will be clear below, why we are going to consider both these cases.

Now, we are ready to write the Ansatz for the supercharges
\footnote{This Ansatz is a direct generalization of
those supercharges for the $SU(2)$ case, which were explicitly
constructed within the superspace approach in \cite{koz}.}
\be\label{SC1}
Q^i=\bp^\alpha \bpsi^i_\alpha- z^\gamma J_{\gamma}{}^\beta h_\beta{}^\alpha \bpsi^i_\alpha+
\psi^{i\,\alpha}f_{\alpha}{}^\beta R_\beta, \quad
{\overline Q}_i=\psi^\alpha_i p_\alpha+ \psi^\alpha_i h_\alpha{}^\beta J_\beta{}^\gamma \bz_\gamma+
\bR^\beta f_\beta{}^\alpha \bpsi_{i\,\alpha}.
\ee
Here, $h_{\alpha}{}^\beta$ and $f_{\alpha}{}^\beta$ are arbitrary, for the time being,
functions depending on the bosonic fields $z^\alpha, \bz_\alpha$ only.
Moreover, due to the explicit $U(n)$ symmetry of our construction, which we are going
to keep unbroken, one may further restrict these functions as
\be\label{fh}
h_{\alpha}{}^\beta=h_1\, \delta_\alpha^\beta+h_2\; \bz_\alpha z^\beta, \quad
f_{\alpha}{}^\beta=f_1\, \delta_\alpha^\beta+f_2 \;\bz_\alpha z^\beta,
\ee
where the scalar functions $h_1, h_2, f_1, f_2$ depend now on $x=z\cdot\bz$ only.

The supercharges \p{SC1} have to obey the standard \Nf Poincar\'{e} superalgebra relations \p{Q1}.
Therefore, the closure of superalgebra is achieved if the following equations on functions in
\p{fh} are satisfied
\bea\label{functions}
\left\{ Q,Q\right\} =0 &\Rightarrow&
\left\{ \begin{array}{l}
 f_1^{'} = - (f_1 h_1 + x f_1 h_2), \quad
f_2^{'} = - (2 f_2 h_1 + f_1 h_2 + 2x f_2 h_2),\\
 h_1^{'} = - (h_1^2 - h_2 + x h_1 h_2),\quad f_2 = - f_1 h_1\,
\end{array} \right. \nn
\\
\left\{ Q^i,{\overline Q}_j\right\} =\im \delta^i_j H &\Rightarrow&
h_2^{'} = \frac{1}{2}\, (A f_1^2 h_1^2 + h_1^3), \quad
h_2 = - \frac{1}{2}\, h_1^2, \quad
A f_1^2 = (2 h_1 - x h_1^2),
\eea
where the derivatives are taken with respect to $x$.

The simplest, almost trivial solution of the equations \p{functions} reads
\be\label{sol1}
f_1=f_2=0,\qquad h_1=\frac{1}{z \cdot \bz},\quad h_2=-\frac{2}{(z \cdot \bz)^2}.
\ee
The functions $h_1,h_2$ in \p{sol1} have a singularity at $(z,\bz) \rightarrow 0$. Moreover, they have no any geometric meaning within $CP^n$ geometry. Thus, without $R,{\bar R}$ terms in the Ansatz \p{SC1} the reasonable interaction can not be constructed.

In contrast, with non-zero $f_1,f_2$ functions the solution of \p{functions} is fixed to be
\bea\label{sol}
&& f_1 =\frac{1}{\sqrt{1+A\, z \cdot \bz}},  \quad f_2=
-\frac{A}{\left( 1+A\, z \cdot \bz\right)\left(1+\sqrt{1+A\, z \cdot \bz}\right)}, \nn \\
&& h_1=\frac{A}{\sqrt{ 1+A\, z \cdot \bz }\left(1+\sqrt{1+A\, z \cdot \bz}\right)}, \quad
h_2=- \frac{1}{2\, \left( 1+A\, z \cdot \bz\right)\left(1+\sqrt{1+A\, z \cdot \bz}\right)^2}.
\eea
Thus, we see that the matrix valued function  $f_{\alpha}{}^\beta$ perfectly coincides
with the vielbeins for the $CP^n$ manifold \p{vb} if we choose $A=1$.
The background gauge field $h_{\alpha}{}^\beta$ is the part of the $U(n)$-connection \p{con} for $CP^n$.
It is worth to note that this  field is identical to the one constructed in
\cite{kn} as the solution of the Bogomol'nyi equation for the Tchrakian's type of self-duality
relations in $U(n)$ gauge theory \cite{T1, T2}.

The last step is to write the Hamiltonian
\bea\label{Ham}
H&=& \left( \bp\, g^{-1}\, p\right) +\left( \bp\, g^{-1}\, h\, J \bz\right) - \left( z\, J\, h \, g^{-1}\,p\right)
-\left(\bR\,f\,g^{-1}\,f\,R\right) -\left(z\, J\, h\,g^{-1}\,h\,J\,\bz\right)\\
&+& i \left ( \frac{1-A}{(1+z \cdot \bar z)(1+A z \cdot \bar z) }
\right) \left ( \left(z\,\bpsi\right)^i\left(\bR\,f\,\bpsi\right)_i -
\left(\psi\,f\,R\right)^i \left( \psi\,\bz\right)_i \right)
- i A\, \left (\psi_i f\, J \,f \bpsi^i \right) \nn\\
&+&\frac{1}{4}\, \left(  g_\mu{}^\alpha g_\rho{}^\sigma  + g_\mu{}^\sigma g_\rho{}^\alpha \right)
\bpsi_{\alpha\, i}\bpsi^i_\sigma\, \psi^{\rho\,j}\psi^\mu_j. \nn
\eea
Here, we used concise notations - all indices in parenthesis are in the proper positions
and they are converted from top-left to down-right, e.g. $\left(\psi\,\bz\right)_i= \psi^\alpha_i\,\bz_\alpha$, etc.

This Hamiltonian commutes with all our supercharges, as it should be.
Its bosonic part (the first line in \p{Ham}) contains the terms describing the interaction
with $U(n)$  background fields and a specific potential term.
The parameter $A$ takes two values $A=\pm 1$, according with the algebra \p{su}.
If we take $A=1$, so the algebra of currents of the internal group is $SU(1,n)$, then the Hamiltonian drastically simplified to be
\bea\label{Ham1}
H_{A=1} &=& \left( \bp\, g^{-1}\, p\right) +\left( \bp\, g^{-1}\, h\, J \bz\right) - \left( z\, J\, h \, g^{-1}\,p\right)
-\left(\bR\,R\right) -\left(z\, J\, h\,g^{-1}\,h\,J\,\bz\right)\\
&-&  \im\, \left (\psi_i f\, J \,f \bpsi^i \right)
+\frac{1}{4}\, \left(  g_\mu{}^\alpha g_\rho{}^\sigma  + g_\mu{}^\sigma g_\rho{}^\alpha \right)
\bpsi_{\alpha\, i}\bpsi^i_\sigma\, \psi^{\rho\,j}\psi^\mu_j. \nn
\eea
Clearly, the $R,\bR$ dependent term in the Hamiltonian \p{Ham1} can be rewritten through the Casimir operator
${\cal K}$ of $SU(1,n)$ algebra
\be\label{Kaz}
{\cal K} = \bR{}^\alpha R_\alpha -\frac{1}{2} J_\alpha{}^\beta J_\beta{}^\alpha +\frac{1}{2(n+1)}J_\alpha{}^\alpha J_\beta{}^\beta.
\ee
Thus, the Hamiltonian depends only on $U(n)$ currents $J_\alpha{}^\beta$ and $SU(1,n)$ Casimir operator \p{Kaz}.

The $U(1)$ gauge potential presented in \p{SC1}, \p{Ham1} has the standard form
\be\label{u1}
{\cal A}_{U(1)} = i \frac{ \dot{z} \bz -z \dot{\bz}}{2\left( 1 + z \cdot \bz\right)}.
\ee
In the simplest case of $CP^1$ we have only this gauge potential in the theory,
while the scalar potential term acquires the form
\footnote{ This is just the example of super-oscillator potential on $CP^n$ manifolds
constructed in \cite{newNers,newNers1}. See also \cite{BellCastNers}.}
\be\label{pot1}
{\cal V}_{CP^1} = - \bR^\alpha R_\alpha -\frac{ z \cdot \bz}{4} J^2.
\ee
Let us remind that we choose the matrix-valued operators $\bR,R,J$ to be anti-hermitian \p{anti}.
Thus, the potential \p{pot1} is positively defined.

Finally, we would like to say a few words about the explicit realization of the isospin groups
$SU(n+1)$ and/or $SU(1,n)$ \p{su}. The common way to involve the isospin variables in
the supersymmetric theories is to introduce the set of semi-dynamical bosonic variables -
harmonics and construct the currents from them (see e.g. \cite{IV} and references therein).
The same strategy could be applied in the present model too.

\section{Conclusion and Discussion}
In the present paper we have constructed a \Nf supersymmetric extension of mechanics describing
the motion of a particle over $CP^n$ manifold in the presence of background $U(n)$ fields.
The gauge potential is proportional to the $U(n)$-connection on $SU(n+1)/U(n)$.
Such a type of background gauge fields has been known for quite a long time in a purely bosonic case \cite{CD}.
What is really nice is that this field appears in our system automatically,
as a result of imposing \Nf supersymmetry. Moreover, in addition to gauge fields
\Nf supersymmetry demands additional potential terms to be present in the Hamiltonian.
In the simplest case of the $CP^1$ system this potential is just a harmonic oscillator one.

One of the most unexpected features of the present model is a strange interplay between the isospin
group which our background gauge fields are coupled to and the form of these fields.
It turns out that the standard $SU(n+1)/U(n)$ $U(n)$-connection appears as a gauge fields
potential only in the case when isospin group is chosen to be $SU(1,n)$.
Alternatively, the choice of the $SU(n+1)$ group for the isospin variables gives rise
to a $U(n)$-connection on the $SU(1,n)/U(n)$ group.
At any rate, both cases are compatible with \Nf supersymmetry.

Another interesting peculiarity of our model is the presence of the isospin variables
on the whole $SU(n+1)$ (or $SU(1,n)$) group, despite the fact that only $U(n)$ background fields
appear in the Hamiltonian. Again, this situation is not new.
The same effect has been noted in the recently constructed \Nf supersymmetric mechanics coupled
to non-Abelian gauge fields \cite{Iv1, Iv2, Iv3, S1, S2, S3}.

One of the possible immediate applications of the constructed model is the analysis of
the role the additional fermionic variables play in the quantum Hall effect on $CP^n$
\cite{Kar1,Nair1,BellCastNers}.
In this respect it could be important that \Nf supersymmetry insists on the simultaneous appearance
of the gauge fields on $U(1)$ and $SU(n)$ with a proper fixed relative coefficient.
The role of the special type of scalar potential which appears due to \Nf supersymmetry also
has to be clarified.

Another interesting possibility to describe \Nf supersymmetric $CP^n$ mechanics is to replace
from the beginning the linear chiral supermultiplets by the nonlinear ones  \cite{IKL1}.
This case is under investigation at present.

\section*{Acknowledgements}
We are indebted to Armen Nersessian, Andriy Shcherbakov and Armen Yeranyan for valuable discussions.

S.K. and A.S. thank the INFN-Laboratori Nazionali di Frascati, where this work was completed, for
warm hospitality.

This work was partly supported by RFBR grants  09-02-01209,
 11-02-90445-Ukr, 11-02-01335, as well as by the ERC
Advanced
Grant no. 226455, \textit{``Supersymmetry, Quantum Gravity and Gauge Fields''%
} (\textit{SUPERFIELDS}).

\end{document}